# Requirements of a Recovery Solution for Failure of Composite Web Services

Hadi Saboohi[1], Sameem Abdul Kareem[2]
[1,2]Department of Artificial Intelligence,
Faculty of Computer Science and Information Technology
University of Malaya (UM), 50603 Kuala Lumpur, Malaysia
[1]saboohi@siswa.um.edu.my

*ABSTRACT*

*Web services are building blocks of interoperable systems. Composing Web services makes the processes capable of doing complex tasks. Composite services may fail during their execution which can be diagnosed by a mediator. The mediator adapts the structure so that the failure is recovered. Moreover, future executions should avoid the situation or organize a strategy to repair the structure with a minimum delay. In this paper the failure reasons of a composite service are reviewed. Furthermore, the requirements of a solution for recovery of a system from a failure are investigated.*

*KEYWORDS*

*Web Service, Composite Web Service, Failure Recovery, Adaptation, Requirement*

## 1. INTRODUCTION

Web processes use services to accomplish users' needs. A single service is usually not able to perform all the required tasks of a business process. The smaller services are integrated to represent a synergy of services. The composite service is executed by a mediator, which stands between service provider and service consumer. The mediator handles the negotiations among the providers and their consumers [1].

The mediation commences from the publication of service specifications. Moreover, the mediator discovers the right service to fulfill the consumers' goal. The composition of a synergy of services occurs if there is no single service available to perform the task. The ultimate responsibility of the mediator is the invocation of either a single or a composite service for the requester and continuous monitoring of the execution [2], [3].

The monitoring ensures a smooth execution of the services such that even if an execution failure happens, the consumer receives the desired result. This smooth execution includes fulfillment of both the required functional properties as well as promising non-functional properties, i.e. quality of service (QoS).

The functional properties include input, output and conditions. The availability of the required input and conditions means the generation of the stated outputs and results by a successful execution of the service.

The non-functional properties are related to the competing properties of different providers, which help the users to choose the particular service among all available functionally equivalent services. The non-functional properties include execution time, cost, and reliability [4] of the service.

In this paper we study the reasons of failure of a composite Web service and the requirements for a solution to its recovery. The major solutions are investigated to show to what extent they have fulfilled the requirements.

The remainder of the paper is structured as follows. In the next section we investigate the failure reasons. Section 3 describes the requirements of a failure recovery solution. Section 4 elaborates the major solutions and their fulfillment of the requirements. Finally Section 5 delivers some conclusion.

## 2. FAILURE REASONS

A successful execution of a Web process featuring several smaller services ties in the well-execution of its services. An execution of a service might be unsuccessful due to several reasons. The reasons are based on the functional and non-functional causes at the top level. The mediator enhances future executions of the composite service by reasoning about the causes of the failure. This enhancement is done by either avoiding the situation that causes the problem or being prepared for a solution to complete the perturbed execution.

A Web service, as a participant in a structure of a composite service may fail due to the following causes [5-9].

### 2.1. Functional Causes

- Malfunctioning of the service: This is usually because of the application level errors.
- Unavailability of the service: The unavailability can be temporary or permanent.
    - The service may disappear permanently. The provider may not provide the service anymore or it may replace the service with a new one.
    - The connecting network has a failure, for example infrastructure breakdown.
    - Host overload. The number of requests is too high that the hosting is unable to serve.
    - User mobility. For example the user changes an accessing network which restricts its external access.

The first unavailability cause is usually permanent and the others are temporary. The temporary causes can be remedied by for example repairing the network or through the introduction of an extra host.
- Software compatibility issues
    - The mismatches among the composed services. For example the changes related to the input and output formats.
    - Malformed response or errors related to serialization/deserialization. For example the changes of negotiating messages.
- The emergence of new requirements: Usually because of the reconfiguration which aims to enhance the fulfillment of needs.
- Changes in the context and the environment. Sometimes even though all the conditions at the provider side are constant, but, the context changes; such as, the accessing device may cause a disorder in getting the desired service.

### 2.2. Non-functional causes

A deviation from a promised quality of service influences the user not to be satisfied with the service. The factors are related to time and cost of the execution which are as follows:
- Response time-out
- Network delay: For example because of network congestion

- Host overload: The provider cannot answer all the requests on time because of unexpectedly:
    – large number of requests for a service
    – large number of invocation of various services of the provider
- Unexpected input
- Unexpected data size

## 3. REQUIREMENTS

It is obviously critical that the mediator must monitor the execution and be aware of the aforementioned causes. It should handle the errors to provide a smooth execution of the Web process. Delegating the error handling to the providers causes interoperability issues.

It is straightforward for the mediator to stop the failed process and generate a new process without using the failed component and start the execution of the new services. However, this enforces a long delay which is not desired by the end-user [10]. The composite service must be repaired and adapted to the new situation.

Arguably, the adaption is categorized as Perfective, Corrective, Adaptive, Preventive, and Extending [11]. Failure recovery is mostly a *corrective* adaptation which removes an undesired or faulty behavior of a system. On the other hand, *adaptive* adaptation is a process which responds to the context, interaction, and requirement changes of the application, and *preventive* adaptation, which prevents from future failures, are also applicable.

In order to propose a solution for the failure recovery of a composite service, the requirements for a process which is able to rectify such a problem is investigated. The requirements are as follows:

- Automation
- Adaptation Probability
- Time Complexity
- Accuracy
- QoS Deviation
- Consideration of World-altering Actions
- Experiments on a Standard Test Collection

In the following sections each item is elaborated.

### 3.1. Automation

The first and foremost requirement is that the adaptation is done automatically and autonomously. The method should minimize the human intervention for the adaptation. Hence, the applications should be able to discover, rank, and compose new services. The semantic descriptions of services [12] are obvious necessities which help to automate the discovery, matchmaking and composition issues.

### 3.2. Adaptation Probability

The adaptation process tries to amend the structure so that its execution can be completed. The probability of the success of such an approach differs. The probability should be high enough so that in most cases the mediator can recover the system from a failure and the system works smoothly even with an unavoidable failure.

## *3.3. Time Complexity*

The adaptation requires some calculations and interactions. The time complexity of such extra processes should be reasonable so that the adaptation is done in a minimum delay. Minimizing the delay isolates the awareness of the user from any likely failure occurrence and its recovery process.

## *3.4. Accuracy*

The adaptation demands for a replacement and an amendment in the primary structure of the services. Some of the smaller services of the composite service must be switched with others. Ideally, the replaced services should have exactly the same functionality or at least be similar practically.

The accuracy is critical since the end-user should get the requested goal. That is to say the goal of the "Adapted Composite Service" should be similar to the goal of the "Original Composite Service".

Semantic Web services ensures an unambiguous description of the Web services such that their discovery and composition occurs automatically and most importantly accurate. Furthermore, the clear distinction of two major groups of the services, i.e. information providing and world-altering services based on the existence of an effect is a major contribution of using semantic Web services for Web processes. Hence, the mediator discovers and composes world-altering services along with the information providing services [13].

## *3.5. QoS Deviation*

The adapted structure of the composite service must be functionally equivalent to the original structure. Additionally, non-functional properties of the composite service must be the same as the promised service to the end-user. Thus, there should be no compromise on the QoS.

## *3.6. Consideration of World-altering Actions*

The approach must contemplate all kinds of services, i.e. information providing and world-altering [12]. Hence, there should be methods to cope with the specific features of world-altering services such as their effects on the real world.

**Table 1. Fulfillment of Requirements**

| Research Paper | Automatic (Human Intervention) | Probability of Adaptation | Time Complexity | Accuracy | QoS Deviation | World-altering Services |
|---|---|---|---|---|---|---|
| **Yu & Lin 2005 [16]** | ✓ | High | Low | High | ✓ | ✗ |
| **Chafle et al. 2006 [17]** | ✓ | High | High | High | ✓ | ✗ |
| **Feng et al. 2007 [18],[19]** | ✓ | High | High | High | ✓ | ✗ |
| **Saboohi et al. 2008 [20]** | ✓ | High | High | High | ✓ | ✓ |
| **Canfora et al. 2008 [21]** | ✓ | High | High | High | ✓ | ✗ |
| **Dai et al. 2009 [22]** | ✓ | High | Low | High | ✓ | ✗ |
| **Lin et al. 2010 [23]** | ✓ | High | Low | High | ✓ | ✗ |
| **Möller & Schuldt 2010 [24]** | ✓ | High | Low | High | ✗ | ✗ |

### *3.7. Experiments on a Standard Test Collection*

Ultimately, there is a strong need for a standard test collection of composite Web services. Current test collections of Web services and semantic Web services do not contain any composite service and they are just sets of single (atomic) services. The standard test collection definitely needs a set of world-altering services to be used for a test on both the information-providing and world-altering services [14], [15].

A failure recovery approach for composite Web services need to be tested on a standard test collection to prove its applicability, accuracy, etc.

## 4. MAJOR SOLUTIONS

According to the aforementioned requirements for an automatic failure recovery method we have investigated the major approaches and the summary is shown in Table 1.
The approaches are briefly described as follows.

### *4.1. Backup Path*

The researchers in [16] proposed two algorithms to adapt a distributed business process. They create a backup path for every service in the execution path. So, there are two paths from a service to the end. The execution will continue through the second path if the first optimal path fails. The approach reconfigures the structure such that the future executions do not use the failed service.

### *4.2. Multiple Stage Adaptation*

The authors of [17] presented a multiple staged approach to the adaptation of Web services. The approach uses either a different instance of a service type in the composition or a different template of services. The recovery process is designed in an incremental staged model.

### *4.3. Multiple QoS Constraints*

Feng et al. [18], [19] presented a composition and a recovery model which satisfies multiple QoS constraints. The method finds another way starting from a preceding service to complete the required task. The new path is found based on maximization and minimization of some QoS values.

### *4.4. Subgraph Replacement*

The authors in [20] introduced an offline-online approach for the recovery based on subgraph replacement. A topmost ranked subgraph of services in a graph of composite service is replaced with another subgraph which has similar functional and non-functional properties.

### *4.5. Rebinding*

Canfora et al. [21] presented a rebinding approach focusing on QoS-aware run-time binding. The slice to be executed is rebound when it is needed. The deviation of QoS is predicted for *case* and *switch* nodes of the graph.

## *4.6. Performance Prediction*

A self-healing solution is presented in [22] using a performance prediction method. The healing includes finding a backup in selection and re-selection in execution. The prediction is particularly for data transmission speed.

## *4.7. Region Reconfiguration*

Kwei-Jay Lin et al. [23] presented a dynamic reconfiguration for the service processes which preserves the initial QoS constraints. The method replaces the failed service together with its neighbors. The region is expanded until a reasonable replacement is found based on QoS constraints.

## *4.8. Dynamic Substitution*

The researchers in [24] proposed a dynamic modification of a composite service at failure time. The approach replaces a semantically equivalent set of services by another set.

## 5. CONCLUSION

There might be many reasons that a composite Web service fails during its execution. The mediator should recover the system from a total failure by adapting the structure. A failure recovery approach of composite Web services requires several major features including its automation, high probability of adaptation, low time complexity, reasonable accuracy, a minimum QoS deviation, and consideration of both information providing and world-altering type of services. We have investigated the major solutions of corrective adaptation based on the requirements.

## REFERENCES

[1] D. Kuropka, P. Tröger, S. Staab, and M. Weske, Eds., *Semantic Service Provisioning*. Berlin: Springer, 2008.
[2] W3C Working Group, "Web Services Glossary," Website, February 2004, http://www.w3.org/TR/ws-gloss/.
[3] T. Erl, *SOA Principles of Service Design (The Prentice Hall Service-Oriented Computing Series from Thomas Erl)*. Upper Saddle River, NJ, USA: Prentice Hall PTR, 2007.
[4] N. Sasikaladevi, L. Arockiam, "Reliability evaluation model for composite web services," *International Journal of Web & Semantic Technology*, vol. 1, no. 2, pp. 16–22, 2010.
[5] E. Di Nitto, C. Ghezzi, A. Metzger, M. Papazoglou, and K. Pohl, "A journey to highly dynamic, self-adaptive service-based applications," *Automated Software Engineering*, vol. 15, no. 3-4, pp. 313–341, December 2008.
[6] K.-J. Lin, J. Zhang, and Y. Zhai, "An efficient approach for service process reconfiguration in SOA with end-to-end QoS constraints," in *IEEE Conference on Commerce and Enterprise Computing (CEC)*. Washington, DC, USA: IEEE Computer Society, July 2009, pp. 146–153.
[7] K. Mahbub and A. Zisman, "Replacement policies for service-based systems," in *The International Conference on Service-Oriented Computing (ICSOC)*, ser. ICSOC/ServiceWave'09. Berlin, Heidelberg: Springer-Verlag, 2009, pp. 345–357.
[8] R. Vaculín, K. Wiesner, and K. Sycara, "Exception handling and recovery of semantic web services," in *Fourth International Conference on Networking and Services (ICNS)*, March 2008, pp. 217–222.
[9] Y. Zhai, J. Zhang, and K.-J. Lin, "SOA middleware support for service process reconfiguration with end-to-end QoS constraints," in *IEEE International Conference on Web Services (ICWS)*, July 2009, pp. 815–822.
[10] Y. Yan, P. Poizat, and L. Zhao, "Repair vs. recomposition for broken service compositions," in *8th International Conferenece on Service-Oriented Computing (ICSOC)*, ser. Lecture Notes in

Computer Science, vol. 6470. San Francisco, CA, USA: Springer, December 2010, pp. 152–166.

[11] R. Kazhamiakin, S. Benbernou, L. Baresi, P. Plebani, M. Uhlig, and O. Barais, "Adaptation of service-based systems," in *Service Research Challenges and Solutions for the Future Internet*, ser. Lecture Notes in Computer Science. Springer Berlin / Heidelberg, 2010, vol. 6500, pp.117–156.

[12] S. A. McIlraith, T. C. Son, and H. Zeng, "Semantic web services," *IEEE Intelligent Systems*, vol. 16, no. 2, pp. 46–53, 2001.

[13] H. Saboohi and S. Abdul Kareem, "World-altering semantic web services discovery and composition techniques - a survey," in *The 7th International Conference on Semantic Web and Web Services (SWWS)*, Las Vegas, Nevada, USA, July 2011, pp. 91–95.

[14] U. Küster and B. König-Ries, "Towards standard test collections for the empirical evaluation of semantic web service approaches," *International Journal of Semantic Computing*, vol. 2, no. 3, pp. 381–402, 2008.

[15] H. Saboohi and S. Abdul Kareem, "A resemblance study of test collections for world-altering semantic web services," in *International Conference on Internet Computing and Web Services (ICICWS) in The International MultiConference of Engineers and Computer Scientists (IMECS)*, vol. I, International Association of Engineers. Hong Kong: Newswood Limited, March 2011, pp. 716–720.

[16] T. Yu and K.-J. Lin, "Adaptive algorithms for finding replacement services in autonomic distributed business processes," in *The 7th International Symposium on Autonomous Decentralized Systems (ISADS)*, April 2005, pp. 427–434.

[17] G. Chafle, K. Dasgupta, A. Kumar, S. Mittal, and B. Srivastava, "Adaptation in web service composition and execution," in *IEEE International Conference on Web Services (ICWS)*. Chicago, USA: IEEE Computer Society, September 2006, pp. 549–557.

[18] X. Feng, H. Wang, Q. Wu, and B. Zhou, "An adaptive algorithm for failure recovery during dynamic service composition," in *Pattern Recognition and Machine Intelligence*, ser. Lecture Notes in Computer Science, A. Ghosh, R. De, and S. Pal, Eds. Springer Berlin / Heidelberg, 2007, vol. 4815, pp. 41–48.

[19] X. Feng, Q. Wu, H. Wang, Y. Ren, and C. Guo, "ZebraX: A model for service composition with multiple QoS constraints," in *Advances in Grid and Pervasive Computing*, ser. Lecture Notes in Computer Science, C. Cérin and K.-C. Li, Eds. Springer Berlin / Heidelberg, 2007, vol. 4459, pp. 614–626.

[20] H. Saboohi, A. Amini, and H. Abolhassani, "Failure recovery of composite semantic web services using subgraph replacement," in *International Conference on Computer and Communication Engineering (ICCCE)*, Kuala Lumpur, Malaysia, May 2008, pp. 489–493.

[21] G. Canfora, M. Di Penta, R. Esposito, and M. L. Villani, "A framework for QoS-aware binding and re-binding of composite web services," *Journal of Systems and Software*, vol. 81, pp. 1754–1769, October 2008.

[22] Y. Dai, L. Yang, and B. Zhang, "QoS-driven self-healing web service composition based on performance prediction," *Journal of Computer Science and Technology*, vol. 24, pp. 250–261, 2009.

[23] K.-J. Lin, J. Zhang, Y. Zhai, and B. Xu, "The design and implementation of service process reconfiguration with end-to-end QoS constraints in SOA," *Service Oriented Computing and Applications (SOCA)*, vol. 4, no. 3, pp. 157–168, June 2010.

[24] T. Möller and H. Schuldt, "OSIRIS Next: Flexible semantic failure handling for composite web service execution," in *Fourth International Conference on Semantic Computing (ICSC)*. Los Alamitos, CA, USA: IEEE Computer Society, September 2010, pp. 212–217.